\documentclass[12pt]{article}
\usepackage{epsfig,amsfonts,amsthm}
\usepackage{amsmath,amssymb}
\newcommand{\be}{\begin{equation}}
\newcommand{\ee}{\end{equation}}
\newcommand{\bea}{\begin{eqnarray}}
\newcommand{\eea}{\end{eqnarray}}

\newcommand{\fdf}[2]{ \phi^\dagger_{#1} \phi_{#2}}

\newcommand{\Z}{\mathbb{Z}}

\def\lsim{\mathrel{\rlap{\lower4pt\hbox{\hskip1pt$\sim$}}
    \raise1pt\hbox{$<$}}}         
\def\gsim{\mathrel{\rlap{\lower4pt\hbox{\hskip1pt$\sim$}}
    \raise1pt\hbox{$>$}}}         

\title{\normalsize \hfill CFTP/13-004 \\[10mm]
\LARGE Geometrical $CP$ violation in the $N$-Higgs-doublet model}
\author{I.~P.~Ivanov$^{1,2}\thanks{E-mail: igor.ivanov@ulg.ac.be}$
\ and
L.~Lavoura$^{3}\thanks{E-mail: balio@cftp.ist.utl.pt}$ 
\\
{\small $^1$ IFPA, Universit\'{e} de Li\`{e}ge,
All\'{e}e du 6 Ao\^{u}t 17, b\^{a}timent B5a, 4000 Li\`{e}ge, Belgium}
\\
{\small $^2$ Sobolev Institute of Mathematics,
Koptyug avenue 4,
630090 Novosibirsk,
Russia}
\\
{\small $^3$ CFTP, Instituto Superior T\'ecnico,
Universidade T\'ecnica de Lisboa, 1049-001 Lisboa, Portugal}
}

\begin{document}

\maketitle

\begin{abstract}
Geometrical $CP$ violation is a particular type of spontaneous $CP$ violation
in which the vacuum expectation values have phases which are calculable,
\textit{i.e.}\ stable against the variation
of the free parameters of the scalar potential.
Although originally suggested
within a specific version of the three-Higgs-doublet model,
it is a generic phenomenon.
We investigate its viability and characteristic features
in models with several Higgs doublets.
Our work contains both general results and illustrative examples.
\end{abstract}

\section{Introduction}

\subsection{Geometrical $CP$ violation}

$CP$ violation~\cite{book} is a hallmark feature
of the electroweak interactions.
Despite a long history of experimental and theoretical exploration,
its fundamental origin remains enigmatic.
Various models of physics beyond the Standard Model (bSM)
try to provide an explanation of that origin.
One attractive idea is that $CP$ is a valid symmetry of the Lagrangian
but gets broken spontaneously by the vacuum state.
In a seminal work,
T.~D.~Lee~\cite{TDLee} has shown that
spontaneous $CP$ violation can take place in the scalar sector
of the two-Higgs-doublet model (2HDM);
later on,
when the existence of three quark generations became known,
it was shown~\cite{Branco1985} that
the phase generated by spontaneous $CP$ violation
in the vacuum state of the 2HDM
actually propagates into the quark mixing matrix.
Since then,
spontaneous $CP$ violation in the 2HDM remained a rich subject, 
with several of its facets uncovered just quite recently~\cite{CPexamples};
a useful review is ref.~\cite{review2011}.

In the 2HDM,
spontaneous $CP$ violation arises in the form of a relative phase between
the vacuum expectation values (VEVs)
of the neutral components of the two Higgs doublets,
despite the fact that all the parameters of the scalar potential
are real.\footnote{A different form of spontaneous $CP$ violation,
in which it occurs in spite of all the VEVs being real,
has been proposed in ref.~\cite{lavoura}.}
This relative phase depends on the numerical values of those parameters.
This fact renders the model less predictive than one might have wished,
since one can tune the parameters of the potential
and thereby produce any phase that one wants;
in addition,
the phase is subject to renormalization.

With larger scalar sectors,
a new form of spontaneous $CP$ violation
known as \textit{geometrical $CP$ violation}\/ can take place.
In it,
the relative phases of the VEVs become {\em calculable},
\textit{i.e.}\ their values do not explicitly depend
on the free parameters of the potential,
rather they arise from its structural properties.\footnote{The
statement that the relative phases of the VEVs in geometrical $CP$ violation
do not depend on the free parameters of the potential is sometimes found.
This statement is somewhat misleading since those parameters
must be such that the $CP$-violating VEV alignment
is \emph{the global minimum} of the potential.
A better definition of calculable phases is that they are
phases which are stable against reasonably large variations
of the free parameters of the potential.}
Besides having a more transparent origin,
such phases are also stable against renormalization. 

Geometrical $CP$ violation was originally discovered~\cite{geometricT}
in the three-Higgs-doublet model (3HDM).
Generically,
the tree-level scalar potential of that model
can be written as a sum of one part
which is invariant under any phase rotations of the doublets,
\be
V_0 = 
\sum_{k=1}^N \left[ - m_k^2 \phi_k^\dagger \phi_k
+ \lambda_k \left( \phi_k^\dagger \phi_k \right)^2 \right]
+ \sum_{1 \le k < l \le N} \left[
\lambda_{kl} \left( \phi_k^\dagger \phi_k \right)
\left( \phi_l^\dagger \phi_l \right)
+ \lambda^\prime_{kl} \left( \phi_k^\dagger \phi_l \right)
\left( \phi_l^\dagger \phi_k \right) \right],
\label{V0}
\ee
with $N=3$,
and a phase-dependent part $V_{\rm phase}$.
In the model considered in ref.~\cite{geometricT}, 
the potential was assumed to be invariant
under the finite group $\Delta(27)$ of Higgs-family transformations
which is generated by the cyclic permutations of the three doublets
and by the rephasing transformation
\be
\phi_2 \to \omega \phi_2, \
\phi_3 \to \omega^2 \phi_3,
\ee
where $\omega = \exp{\left( i\, 2 \pi / 3 \right)}$.
With this $\Delta(27)$ symmetry,
the scalar potential is very restricted
and only one phase-sensitive term survives:
\be
V_{\rm phase} = \lambda \left[
\left( \fdf{1}{2} \right) \left( \fdf{1}{3} \right) + 
\left( \fdf{2}{3} \right) \left( \fdf{2}{1} \right) + 
\left( \fdf{3}{1} \right) \left( \fdf{3}{2} \right) + 
{\rm H.c.} \right].
\ee
It was noted in ref.~\cite{geometricT} that,
for positive $\lambda$,
the alignment of the VEVs at the global minimum is either
\be
\left(
\left\langle \phi_1^0 \right\rangle_ 0,\,
\left\langle \phi_2^0 \right\rangle_ 0,\,
\left\langle \phi_3^0 \right\rangle_ 0
\right) = u \left( 1,\, 1,\, \omega \right),
\label{vev3HDM}
\ee
with positive real $u$,
or other alignments related to the one in eq.~(\ref{vev3HDM})
by complex conjugation and/or
by a transformation belonging to the symmetry group $\Delta(27)$.

The subject of geometrical $CP$ violation lay dormant for decades but,
recently,
interest in it has revived.
Continuing the analysis of the original $\Delta(27)$-symmetric 3HDM potential,
the authors of ref.~\cite{geometricCPnonR} have shown that
phases remain calculable even when
effective non-renormalizable scalar interactions
of arbitrarily high order are present in that model.
In ref.~\cite{geometricCP} the idea of geometrical $CP$ violation
was extended to models with several Higgs doublets.
On a more mathematical side,
a recent work~\cite{minimization2012} has argued that calculable phases,
and in general any stable kind of alignment of VEVs,
may arise in models whose orbit space has sharp corners;
this remark provides an additional perspective
on the meaning of the word `geometrical'
in the phrase `geometrical $CP$ violation'.
Finally,
in ref.~\cite{Holthausen:2012dk} the question of when calculable phases
lead to genuine $CP$ violation was recast in the form
of a distinction between inner and outer automorphisms
of the Higgs-family-symmetry group of the potential;
that work,
however,
referred only to the original 3HDM example of geometrical $CP$ violation 
in ref.~\cite{geometricT}.

\subsection{Goals and scope of this work}

In this work,
we report on the results of a
study of geometrical $CP$ violation
in multi-Higgs-doublet models that we have undertaken.

Our work differs from ref.~\cite{geometricCP} in several aspects.
Firstly,
after deriving calculable phases of VEVs,
one has to check that they cannot be eliminated
by a rephasing symmetry of the potential.
This check becomes automatic
once we know the rephasing symmetry group of the given potential.
So,
we start our investigation by developing a theory
for finding it for any $N$-Higgs-doublet model (NHDM) potential.
Secondly,
we consider only renormalizable potentials,
since they offer already quite a rich spectrum of possibilities.
The inclusion of non-renormalizable terms goes beyond the scope of this work,
but they can also be analyzed 
using the methods developed here.
Thirdly,
we limit the variety of cases by requiring that
the phase-sensitive part of the potential
contains only one cyclic chain of terms,
which is protected by a specific rephasing symmetry group
not shared with any other possible terms.
We call such a symmetry `stable'
because it guarantees that
no other renormalizable terms in the potential
are generated upon renormalization.
Finding out which rephasing symmetry groups are stable
and what are the potentials that they stabilize
is another task that we solve in this paper.
In section~\ref{section-beyond} we shall
also comment
on the consequences of lifting this requirement.

In order to ensure that no additional terms 
arise in the potential upon renormalization,
one usually imposes a symmetry on the Lagrangian.
In the particular case of geometrical $CP$ violation,
it is known~\cite{geometricT} that
one {\em must} impose an additional symmetry beyond just rephasing,
because otherwise one would get only trivial calculable phases $0$ or $\pi$
among the VEVs.
In the original work~\cite{geometricT},
this was formulated as a requirement
of a non-Abelian symmetry group of the potential,
but this formulation is ambiguous because not every Abelian symmetry group
of a NHDM potential
can be represented by rephasing transformations~\cite{abelianNHDM}.

Specifically,
we consider a NHDM whose Higgs potential is invariant under the $\Z_N$ group
of cyclic permutations of the doublets
\be
\phi_1 \to \phi_2 \to \dots \to \phi_N \to \phi_1,
\label{cyclic}
\ee
and  also under the $CP$ transformation
$\phi_k \left( t, \vec x \right) \to \phi_k^\ast \left( t, - \vec x \right)$
for $k = 1, 2, \ldots, N$.
These invariances significantly restrict the form of $V_0$
but still leave much freedom in $V_{\rm phase}$.
Indeed,
$V_{\rm phase}$ may contain both quadratic terms of the form
\be
V_{k} = \lambda_k \left( \fdf{1}{k} + \fdf{2}{k+1}
+ \dots + \fdf{N}{k-1} + {\rm H.c.} \right)
\label{terms-Vk}
\ee
and quartic terms of the form
\bea
V_{klm} &=& \lambda_{klm} \left[
\left( \fdf{1}{k} \right) \left( \fdf{l}{m} \right)
+ \left( \fdf{2}{k+1} \right) \left( \fdf{l+1}{m+1} \right)
\right. \nonumber \\ & & \left.
+ \dots + \left( \fdf{N}{k-1} \right) \left( \fdf{l-1}{m-1} \right)
+ {\rm H.c.} \right].
\label{terms-Vp}
\eea
In eqs.~(\ref{terms-Vk}) and~(\ref{terms-Vp}),
and below,
the subscripts are always understood to be modulo $N$.
Since the $\Z_N$ cyclic symmetry~(\ref{cyclic})
will always be present
in our examples for $V_{\rm phase}$,
we shall from now on,
for the sake of brevity,
write down only the first term followed by the expression
``$+\ {\rm cyclic}$.''

The analysis of spontaneous $CP$ violation
in the NHDM with a generic collection of terms $V_k$ and $V_{klm}$
is a complicated task which goes much beyond the scope of this paper.
Here,
we just want to present some simple examples
realizing the idea of geometrical $CP$ violation for $N > 3$.
For this purpose,
we impose an additional invariance of the potential $V_{\rm phase}$
under a certain rephasing symmetry group $G$;
that additional invariance selects {\em only one}
set of phase-sensitive terms out of all the $V_k$ and $V_{klm}$.
We call the rephasing symmetry $G$ {\it stable}\/
if it guarantees that $V_{\rm phase}$ contains
{\em only one cyclic set of renormalizable terms} compatible with it,
so that no additional symmetric terms may appear upon renormalization.

The criterion of stability immediately allows us
to disconsider all the terms $V_k$ in eq.~(\ref{terms-Vk}).
Indeed,
if $V_k$ were invariant under a certain $G$, 
then $V_{k1k}$ would also be invariant,
and this disagrees with the definition of stable symmetry.
Thus,
in this paper we only have to deal with quartic terms $V_{klm}$.

Finding out stable rephasing symmetries is our first task
and is solved in section~\ref{sec:stable}.
When one such symmetry is found,
we then need to minimize the potential
and check whether geometrical $CP$ violation
may indeed take place in the global minimum.

Minimization of a NHDM potential with several terms
is a highly non-trivial problem
and we do not attempt to solve it in full generality.
We merely argue that
it is always possible to find a region in the space of parameters
such that the VEVs are of the form of pure phases,
\textit{viz.}\ $\left\langle \phi_k^0 \right\rangle_0
= u \exp{\left( i \theta_k \right)}$
with some positive real $u$ which is the same for all $k$.
We then need to find the values of the $\theta_k$ at the global minimum.
If those phases are non-trivial and if,
moreover,
they cannot be all rotated away through a symmetry of the potential,
then we have spontaneous $CP$ violation.
In order to be able to call that $CP$ violation \emph{geometrical},
the phases $\theta_k$ must be calculable,
\textit{i.e.}\ they cannot explicitly depend
on the parameters of the potential.

\section{Stable symmetries}
\label{sec:stable}

\subsection{The rephasing symmetry group and the Smith normal form}
\label{subsection-smith}

The first question that one faces
is how to determine the rephasing symmetry group of a given $V_{klm}$.
This question can be answered in its full generality
by using the methods developed in ref.~\cite{abelianNHDM}.
(See also refs.~\cite{schieren},
where a similar construction has been used.)
Here we slightly adapt those methods to our problem.
Some basic elements of the machinery that we are going to expose in this section
were also used in ref.~\cite{geometricCP}.

Let us consider the full rephasing symmetry group $[U(1)]^N$,
in which the $k$-th $U(1)$ factor
stands for phase rotations of $k$-th Higgs doublet,
$\phi_k \to \phi_k \exp{\left( i \alpha_k \right)}$.
Consider one particular quartic term in $V_{klm}$,
for example the first term $(\fdf{1}{k})(\fdf{l}{m})$.
Upon the action of $[U(1)]^N$,
this term acquires a phase shift
\be
\xi_1 = -\alpha_1 + \alpha_k - \alpha_l + \alpha_m
= \sum_{j=1}^N d_{1j}\alpha_j.
\ee
This term is invariant if $\xi_1 = 2 \pi n_1$ with some integer $n_1$.
Similar conditions,
\textit{viz.}\ $\xi_k = 2 \pi n_k$ with integer $n_k$ for all $k \le N$,
must hold for all the quartic terms in $V_{klm}$.
We thus obtain the system of equations
\be
\sum_{j=1}^N d_{kj} \alpha_j = 2 \pi n_k,
\label{dij-equation}
\ee 
which is expressed via an $N\times N$ matrix of integers
$D = \left[ d_{kj} \right]$.
The entries of $D$ are either $0$,
$\pm 1$,
or $\pm 2$,
and its rows may only be---up to the overall sign
and to cyclic permutations---of one of the three types
\begin{subequations}
\label{allowed-rows}
\bea
& \left( \ldots, 2, \ldots, -2 ,\ldots \right), & \\
& \left( \ldots, 2, \ldots, -1, \ldots, -1, \ldots \right), &
\label{ucity} \\
& \left( \ldots, 1, \ldots, -1 , \ldots, \pm 1, \ldots, \mp 1, \ldots
\right), &
\label{ivuty}
\eea
\end{subequations}
where the dots indicate zeros.
For example,
for
\be
V_{213} = \lambda_{213} \left[
\left( \fdf{1}{2} \right) \left( \fdf{1}{3} \right)
+ {\rm cyclic} + {\rm H.c.} \right],
\label{V213}
\ee
we have
\be
D = \left(\begin{array}{cccccccc}
2 & -1 & -1 & 0 & \cdots & 0 & 0 & 0 \\ 
0 & 2 & -1 & -1 & \cdots & 0 & 0 & 0 \\
\vdots & \vdots & \vdots & \vdots & & \vdots & \vdots & \vdots \\ 
0 & 0 & 0 & 0 & \cdots & 2 & -1 & -1 \\ 
-1 & 0 & 0 & 0 & \cdots & 0 & 2 & -1 \\
-1 & -1 & 0 & 0 & \cdots & 0 & 0 & 2  
\end{array}\right),
\label{aaa}
\ee 
\textit{i.e.},
the rows are of type~(\ref{ucity}).

The solutions of eqs.~(\ref{dij-equation})
form the rephasing symmetry group $G$ of $V_{klm}$.
There is a systematic procedure for finding out this group~\cite{abelianNHDM}.
It is known that by performing elementary transformations
of the following types:
\begin{description}
\item adding one column or one row to another column or row of $D$,
\item flipping the sign of one column or one row of $D$,
\item permuting columns or rows of $D$,
\end{description}
one can bring $D$ to its \textit{Smith normal form},
which is a diagonal matrix
\be
{\rm diag} \left( d_1,\, d_2\, \ldots,\, d_r,\, 0,\, \ldots,\, 0 \right)
\label{smith}
\ee
of positive integers $d_k$
such that each $d_k$ is a divisor of $d_{k+1}$.
In eq.~(\ref{smith}),
$r=\mathrm{rank}\ D$.
Since the elementary transformations described above
preserve the structure of eqs.~(\ref{dij-equation}),
we thus obtain a system of uncoupled equations
\be
d_j \tilde \alpha_j = 2 \pi \tilde n_j,
\ee
for $j = 1,\,\dots,\, r$,
where the $\tilde \alpha_j$ are linear combinations
of the initial $\alpha_k$,
and the $\tilde n_j$ are integers.
The rephasing symmetry group of $V_{klm}$ is then
\be
G = \Z_{d_1} \times \Z_{d_2} \times \cdots \times \Z_{d_r}
\times \left[ U(1) \right]^{N-r},
\ee
where $\Z_1$ is the trivial group.
This result solves in principle the problem of finding out
the rephasing symmetry group of any $V_{klm}$.

Some information on $G$ can be extracted even without explicitly computing
the Smith normal form of $D$.
Since all the elementary steps used in the computation
of the Smith normal form preserve the absolute value of $\det{D}$,
one concludes that either $\left| \det{D} \right| = \prod_{k=1}^N d_k$ if $r=N$
or $\det{D} = 0$ if $r<N$.
In the case of the scalar potential of a NHDM,
$V_{klm}$ is always invariant under one and only one $U(1)$ group,
which is the gauge group of overall phase rotations
$\phi_k \to \phi_k \exp{\left( i \alpha \right)}$;
no other $U(1)$ symmetry group may exist
lest Goldstone bosons arise upon spontaneous symmetry breaking.
We therefore conclude that the matrix $D$ is always singular
with $r=N-1$.
The invariance under the $U(1)$ gauge group does not represent
any interesting property of the potential.
We may remove this uninteresting rephasing symmetry
by considering the $(N-1) \times (N-1)$ matrix $\tilde D$
which is obtained from $D$ by deleting a single column and a single row of $D$.
(These may be \emph{any} row and \emph{any} column.)
The remaining part of the rephasing symmetry group,
$G/U(1)$,
is \emph{finite} and has order
\be
\left| G / U(1) \right| = \left| \det{\tilde D} \right|.
\ee
Now,
if,
in particular,
it happens that $| \det{\tilde D} |$ is a number
whose prime decomposition involves only primes with power one,
\textit{i.e.}\ $|\det{\tilde D}| = p_1 p_2 \cdots p_s$,
then the rephasing symmetry group is uniquely determined to be
\be
G/U(1) = \Z_{\left| \det{\tilde D} \right|},
\ee
since in that case
$\Z_{p_1} \times \Z_{p_2} \times \cdots \times \Z_{p_s}$
is the same as $\Z_{\left| \det{\tilde D} \right|}$.

In other cases,
this simple procedure is not sufficient to find the rephasing symmetry group
and one needs to actually compute the Smith normal form.
Finding the Smith normal form is easily algorithmizable 
and there exist computer-algebra packages \cite{packages}
which do it for any input matrix of integers;
these packages are adequate for practical purposes
and for any reasonable number of doublets.

To keep the notation short,
we shall from now on suppress
the $U(1)$ factor in the rephasing symmetry group $G$,
implicitly assuming that it is present everywhere.

\subsection{The rephasing symmetry group of $V_{213}$ with generic $N$}
\label{subsection-certain-potentials}

Consider the prototypical potential $V_{213}$ in eq.~(\ref{V213})
for a generic number of doublets $N$.
Let us suppose that $G / U(1)$
keeps the first doublet $\phi_1$ invariant
and that it transforms the second doublet $\phi_2$
with a phase factor $\sigma$.
We furthermore suppose that the $k$-th doublet 
is transformed as $\phi_k \to \sigma^{s_k} \phi_k$,
with $s_1 = 0$ and $s_2 = 1$ by construction.
Using the form of the potential,
which contains terms $(\phi_{k-2}^\dagger \phi_{k-1}) (\phi_{k-2}^\dagger \phi_k)$,
one can derive a recurrence relation for the $s_k$,
\be
s_k = 2 s_{k-2} - s_{k-1}.
\ee
Solving this recurrence relation via the \textit{Ansatz}\/ $s_k = a^k$,
we obtain that $a$ may be either $1$ or $-2$.
Inputting the initial conditions $s_1 = 0$ and $s_2 = 1$,
one finally obtains
\be
s_k = \frac{1- \left( -2 \right)^{k-1}}{3}\,,
\ee
which gives the sequence $0, 1, -1, 3, -5, 11, -21, 43, -85, 171, \ldots$.

The fact that the rephasing transformation
is generated by a single factor $\sigma$
means that the group $G$ is in this case of the form $\Z_q$.
In order to find the value of $q$,
we note that the doublet $\phi_{N+1}$ is identical with $\phi_1$,
and therefore must transform with the same phase factor $1$,
\textit{viz.}\ $\sigma^{s_{N+1}} = 1$.
Therefore
the rephasing symmetry group of $V_{213}$ is 
\be
G_{213} = \Z_{|s_{N+1}|}.
\ee

\subsection{Stable symmetries}\label{section-stable}

The next question that we ask is whether $V_{klm}$
has a \emph{stable} rephasing symmetry group,
\textit{i.e.}\ a rephasing symmetry group that allows {\em only} $V_{klm}$
in $V_\mathrm{phase}$.

We firstly note
that $V_{klm}$ and $V_{mlk}$ transform in the same way
under a rephasing transformation.
Therefore,
they have the same rephasing symmetry group.
Since a stable symmetry must allow only one cyclic set of terms,
we conclude that $V_{klm}$ must be identical with $V_{mlk}$.
This is possible only when $l=1$ or
(which is equivalent)
when $k=m$.
As a consequence of this criterion,
from now on we shall consider only sets of terms of the form $V_{k1m}$,
\textit{i.e.}\ with $l=1$.

Next,
consider the matrix $D$ for some $V_{k1m}$.
If it turns out that by summing or subtracting rows of $D$
one can obtain another row---distinct from all the rows of $D$---of
one of the types~(\ref{allowed-rows}), 
then the symmetry is not stable.
If this is impossible,
then $G$ is stable.

We illustrate this criterion by considering $V_{213}$ for $N=4$.
We know from the previous subsection that then $G = \Z_5$. 
The matrix $D$ is
\be
D = \left(\begin{array}{cccc}
2 & -1 & -1 & 0 \\ 
0 & 2 & -1 & -1 \\
-1 & 0 & 2 & -1 \\
-1 & -1 & 0 & 2  
\end{array} \right).
\label{ad213}
\ee 
The sum of the first and third rows of this matrix gives
$(1,\, -1,\, 1,\, -1)$,
which is of the type~(\ref{ivuty}) and,
thus,
allowed.
This means that the terms
$(\fdf{1}{2})(\fdf{1}{3})$ and $(\fdf{3}{4})(\fdf{3}{1})$,
which are present in $V_{213}$ for $N=4$,
generate another quartic term $(\fdf{1}{2})(\fdf{3}{4})$,
which is also invariant under the same $\Z_5$ but is absent in $V_{213}$. 
Therefore,
$V_{213}$ is not protected by a stable $G$ for $N=4$.
In fact,
it can be shown that
{\em in the case $N=4$ there is no stable $G$ for any $V_\mathrm{phase}$}.

In the case $N=5$,
on the other hand,
the sum of the first and third rows of $D$ yields $(2,\, -1,\, 1,\, -1,\, -1)$,
which is not an allowed row.
In this case the terms
$(\fdf{1}{2})(\fdf{1}{3})$ and $(\fdf{3}{4})(\fdf{3}{5})$
generate only
non-renormalizable effective terms
with six doublets,
but no other renormalizable terms.
One may check that no other combination of rows leads in this case
to a new allowed row.
This means that for $N=5$ the potential $V_{213}$ 
is protected by a stable symmetry,
which from the previous subsection we know to be $\Z_{11}$.

\subsection{Stable symmetries for small $N$}

We apply here the methods described above to list all the potentials $V_{klm}$
invariant under stable symmetries for small numbers of doublets.

For $N=3$ there is only $V_{213}$,
which is invariant under a stable symmetry $\Z_3$
and was studied in the seminal paper of ref.~\cite{geometricT}.

For $N=4$,
as we said above,
there is no $V_{klm}$ which can be protected by a stable symmetry group.

For $N=5$,
only
\be
V_{213}\,,\quad V_{315}\,,\quad V_{412}\,,\quad V_{514}
\label{stable5}
\ee
are protected by a stable symmetry,
which is $\Z_{11}$ for all the potentials in~(\ref{stable5})
but with different generators.
Generically,
the potentials in~(\ref{stable5})
can be written as $V_{1+k,\, 1,\, 1+2k}$ with $k=1,2,3,4$
(we remind the reader that all subscripts are meant modulo $5$ in this case).
In each of these four cases,
the generators acts on the doublets in the following way:
\be
\phi_1 \to \phi_1, \quad 
\phi_{1+k} \to \sigma \phi_{1+k}, \quad
\phi_{1+2k} \to \sigma^{10} \phi_{1+2k}, \quad
\phi_{1+3k} \to \sigma^{3} \phi_{1+3k}, \quad
\phi_{1+4k} \to \sigma^{6} \phi_{1+4k},
\label{sym5Vk}
\ee
where $\sigma^{11}=1$.

We do not consider the case $N=6$ in detail here because,
as we shall see later,
there is no geometrical $CP$ violation for stable symmetries
in the case of even $N$.

For $N=7$ there are two distinct sets of potentials,
with six $V_{klm}$ in each set.
The generic form of each set and its respective symmetry group are
\be
V_{1+k,\, 1,\, 1+2k} \ \  \mbox{with} \ \ G = \Z_{43}
\qquad\mbox{and}\qquad
V_{1+k,\, 1,\, 1+3k} \ \ \mbox{with} \ \ G = \Z_{29},
\label{stable7}
\ee
for $k=1,2,3,4,5,6$.
Again,
within each set of $V_{klm}$,
the generators realizing the symmetry group
are different for different potentials.
For the second set in~(\ref{stable7}),
the symmetry group $G$ is
\be
\begin{array}{lclclcl}
\phi_1 \to \phi_1, & & 
\phi_{1+k} \to \rho \phi_{1+k}, & &
\phi_{1+2k} \to \rho^8 \phi_{1+2k}, & &
\phi_{1+3k} \to \rho^{28} \phi_{1+3k},
\\*[2mm]
\phi_{1+4k} \to \rho^{23} \phi_{1+4k}, & &
\phi_{1+5k} \to \rho^{17} \phi_{1+5k}, & &
\phi_{1+6k} \to \rho^4 \phi_{1+6k}, &
\end{array}
\label{rho}
\ee
where $\rho^{29} = 1$.

\section{Calculating the phases}

\subsection{Minimizing the potential}

The simplest form of geometrical $CP$ violation
is characterized by VEVs which differ only by phases:
$\left\langle \phi_k^0 \right\rangle_0 = u e^{i\theta_k}$.
Before calculating the phases $\theta_k$,
we need to prove that the potentials of the type that we consider indeed allow,
without finetuning,
for a global minimum with equal
$\left| \left\langle \phi_k^0 \right\rangle_0 \right|^2 = u^2$
for all $k = 1, \ldots, N$.

Let us first consider only the potential $V_0$.
We assume that its the minimum is neutral,
\textit{i.e.}\ that it does not violate electric-charge conservation.
Substituting the VEVs in $V_0$,
we rewrite it in terms of $u_k^2 \equiv
\left| \left\langle \phi_k^0 \right\rangle_0 \right|^2$.
The expression must respect the cyclic symmetry
and therefore it must be of the form
\be
\left\langle V_0 \right\rangle_0 =
- m_0^2 \sum_{k=1}^N u_k^2 + \frac{1}{2} \sum_{k,l=1}^N \Lambda_{kl} u_k^2 u_l^2.
\label{V0min}
\ee
The real and positive-definite matrix $\Lambda_{kl}$
is symmetric and circulant:
\be
\Lambda_{kl} = \Lambda_{lk}
\quad \mbox{and}\quad
\Lambda_{k+1,l+1} = \Lambda_{kl},
\quad \forall\, k,l = 1, \ldots, N.
\ee
Each row of $\Lambda_{kl}$ is a cyclic permutation of the previous row,
and therefore the sum of all elements in a single row,
$\Lambda \equiv \sum_l \Lambda_{kl}$, 
is the same for all rows.
Equation~(\ref{V0min}) can then be rewritten as
\be
\left\langle V_0 \right\rangle_0 =
\frac{1}{2} \sum_{k,l=1}^N
\Lambda_{kl} \left( u_k^2 - u_0^2 \right)
\left( u_l^2 - u_0^2 \right)
- \frac{u_0^4 N \Lambda}{2},
\quad \mbox{where} \ u_0^2 = \frac{m_0^2}{\Lambda}.
\label{V0min2}
\ee
It then becomes evident that the minimum of $V_0$ occurs,
when $m_0^2$ is positive, 
at $u_k^2 = u_0^2$,
\textit{viz.}\ the minimum respects the cyclic symmetry $\Z_N$.

Next we need to prove that
the inclusion of the phase-sensitive terms in $V_{k1m}$
does not spoil this cyclic symmetry
and only amounts to a shift $u_0 \to u$.
Upon substitution of the VEVs
$\left\langle \phi_k^0 \right\rangle_0 = u_k e^{i \theta_k}$,
$V_{k1m}$ becomes
\be
\left\langle V_{k1m} \right\rangle_0 =
2 \lambda_{k1m}
\left( u_1^2 u_k u_m \cos{\psi_1} + \mathrm{cyclic} \right),
\quad \mbox{where} \ \psi_k = \sum_{l=1}^N d_{kl}\theta_l,
\ee
with the same matrix $D$ as in section~\ref{subsection-smith}.
In order that $u_0$ evolves smoothly to $u$
while keeping all the $u_j$ equal,
we need to require the partial derivatives
$\partial \left\langle V_{k1m} \right\rangle_0 / \partial u_j$
to be equal for all $j$.
Computing those partial derivatives,
we observe that we need to require that the quantities
$c_j = 2 \cos{\psi_j} + \cos{\psi_{j+k-1}} + \cos{\psi_{j+m-1}}$
are all equal at the minimum.
This happens only when all the $\cos{\psi_j}$ are equal.
Later we will see that this condition is indeed verified at the global minimum,
which confirms that the \textit{Ansatz}\/
$\left\langle \phi_k^0 \right\rangle_0 = u e^{i\theta_k}$ is valid.

\subsection{Calculating the phases: general theory}
\label{subsection-calculating-phases}

In this section we consider $V_{k1m}$ and ask what are the phases of the VEVs,
\textit{viz.}\ the $\theta_k$,
that will minimize it.
At the minimum,
the expectation value of $V_{k1m}$ is $2 \lambda_{k1m} u^4 J$,
where
\be
J = \cos{\psi_1} + \cos{\psi_2}
+ \cdots + \cos{\psi_N}.
\label{vk1m-min}
\ee
Obviously,
the stability points of $J$ are points where
\be
\frac{\partial J}{\partial \theta_k} =
- \sum_{j=1}^N \sin{\psi_j}\, d_{jk} =
- \sum_{j=1}^N \left( D^T \right)_{kj} \sin{\psi_j} = 0
\label{vurio}
\ee
for all $k = 1, 2, \ldots, N$.

We have seen before that,
because of the $U(1)$ gauge symmetry,
each term in $V_{klm}$ has an equal number of $\phi$'s and $\phi^\dagger$'s,
\textit{viz.}\ the sum of all the columns of $D$ vanishes identically.
This means that $D$ has a right-eigenvector
$\left( 1,\, 1,\, \ldots,\, 1 \right)^T$ corresponding to the eigenvalue $0$.
Now,
$D$ is a \emph{circulant} matrix,
because of the invariance under~(\ref{cyclic}),
which means that each row of $D$ is equal to the previous row
moved one step to the right.
As a consequence,
the sum of the $D$-matrix elements
over any column of $D$ is equal for all the columns
and it is equal to the sum of the $D$-matrix elements over any row of $D$,
\textit{viz.}\ to zero.
Thus,
the sum of all the rows of $D$ vanishes identically.
This means that $D$ has a left-eigenvector
$\left( 1,\, 1,\, \ldots,\, 1 \right)$ corresponding to the eigenvalue $0$,
or,
equivalently,
that $D^T$ has a right-eigenvector
$\left( 1,\, 1,\, \ldots,\, 1 \right)^T$ corresponding to the eigenvalue $0$.
Therefore,
the extremum points of $J$ given by eqs.~(\ref{vurio}) have
\be
\sin{\psi_1} = \sin{\psi_2} = \cdots = \sin{\psi_N}.
\label{psi-i1}
\ee
It must moreover be kept in mind that,
since the sum of all the rows of $D$ is identically zero,
\be
\sum_{j=1}^N \psi_j = 0\ \mbox{mod}\ 2 \pi.
\label{psi-i2}
\ee
Equations~(\ref{psi-i1}) and~(\ref{psi-i2}) are the ones
which yield the values of the $\psi_j$ at the minimum.

Once the solutions of eqs.~(\ref{psi-i1}) and~(\ref{psi-i2})
for the $\psi_j$ are found,
the phases $\theta_k$ must be obtained from $\psi_j = \sum_k d_{jk} \theta_k$.
The matrix $D$ has determinant zero and
therefore we cannot invert this matrix equation.
Rather we may,
without loss of generality,
set $\theta_1 = 0$ and then define the $(N-1) \times (N-1)$ matrix $\bar D$
which is identical to $D$ with its first row and its first column both removed.
We then have
\be
\psi_j = \sum_{k=2}^N \left( \bar D \right)_{jk} \theta_k
\quad \mbox{for} \ j = 2, \ldots, N,
\label{eqn-theta-i}
\ee
which may be inverted:
\be
\theta_k = \sum_{j=2}^N \left( \bar D^{-1} \right)_{kj} \psi_j
\quad \mbox{for} \ k = 2,\ldots,N.
\label{solution-theta-i}
\ee
Equation~(\ref{solution-theta-i})
is a non-homogeneous matrix equation for the $\theta_k$.
Its general solution can be written as one particular solution
plus the general solution of the corresponding homogeneous equation,
which is
\be
\theta_{k,0} = \sum_{j=2}^N \left( \bar D^{-1} \right)_{kj}
\left( 2 \pi n_j \right),
\label{homogeneous}
\ee
with integer $n_j$.
Equation~(\ref{homogeneous}) has exactly the same form
as eq.~(\ref{dij-equation})
and we already know that the set of its solutions
is the rephasing symmetry group of the potential.

When $\lambda_{k1m} < 0$ the potential is minimized by maximizing $J$.
This is simply done by setting all the $\psi_j = 2 \pi n_j$
with integer $n_j$.
Then eq.~(\ref{solution-theta-i})
becomes identical with the homogeneous eq.~(\ref{homogeneous}).
This means that the $\theta_k$ that we will get are essentially trivial,
\textit{i.e.}\ $\theta_k = 0$ up to a transformation
of the rephasing symmetry group $G$ of $V_{k1m}$.
Hence,
there is no spontaneous $CP$ violation when $\lambda_{k1m} < 0$.

In the opposite case $\lambda_{k1m} > 0$,
we need to minimize $J$ instead.
Now the analysis depends on whether $N$ is even or odd.
In the case of even $N$ the relevant solution
of eqs.~(\ref{psi-i1}) and~(\ref{psi-i2}) is
$\psi_j = \pi$ for all $j = 1, 2, \ldots, N$.
This is a $CP$-conserving situation.
Indeed,
suppose that some non-trivial phases $\theta_k$ solve eqs.~(\ref{eqn-theta-i})
with all the $\psi_j = \pi$.
Then,
the phases $-\theta_k$ also solve the same equations.
Even if the two alignments $\theta_k$ and $- \theta_k$ do not coincide,
they differ by a solution to the
homogeneous eq.~(\ref{homogeneous}),
\textit{i.e.}\ they are linked by a rephasing symmetry transformation.
We then obtain that
\be
\left\langle \phi_k^0 \right\rangle_0^\ast
= \sum_{j=1}^N U_{kj} \left\langle \phi_j^0 \right\rangle_0,
\ee
where $U$ is a unitary symmetry of the potential.
This means that the vacuum is invariant under a
(generalized) $CP$ transformation
which is a symmetry of the Lagrangian.
We have thus found that, 
in cases with a stable rephasing symmetry,
\emph{there can be no geometrical $CP$ breaking if $N$ is even}.

In the case of odd $N$,
eqs.~(\ref{psi-i1}) and~(\ref{psi-i2}) have solutions
\be
\psi_j = \psi_0 = \pi \pm \frac{\pi}{N}\,,
\label{psisol}
\ee
which are $CP$-conjugate of each other
and yield $J = - N \cos{\left( \pi / N \right)}$.
Note that this result is insensitive to the exact form of the potential,
namely to the values of $k$ and $m$ in $V_{k1m}$.

Given the solution~(\ref{psisol}) for the $\psi_j$,
we must construct the particular solution
of eq.~(\ref{solution-theta-i}) through
\be
\theta_k = \psi_0 \sum_{j=2}^N \left( \bar D^{-1} \right)_{kj}
\quad \mbox{for} \ k = 2,\ldots,N.
\label{jbhur}
\ee
This particular solution depends on the structure of the potential,
\textit{viz.}\ of the values of $k$,
$m$,
and $N$.
However,
often there exists another particular solution of eq.~(\ref{solution-theta-i}),
which has the simple form
\be
\theta_k = \eta_k = \frac{2\pi}{N} \left( k-1 \right)
\quad \mbox{for}\ k = 1, 2, \ldots, N,
\label{very-particular}
\ee
satisfying our \textit{Ansatz}\/ $\theta_1 = 0$.
We stress that the existence of the solution~(\ref{very-particular})
should always be explicitly verified in each particular case.

\subsection{First example: $V_{213}$ for $N=3$}

We first consider the case of the potential
\be
V_\mathrm{phase} = \lambda_{213} \left[
\left( \phi_1^\dagger \phi_2 \right) \left( \phi_1^\dagger \phi_3 \right)
+ \left( \phi_2^\dagger \phi_3 \right) \left( \phi_2^\dagger \phi_1 \right)
+ \left( \phi_3^\dagger \phi_1 \right) \left( \phi_3^\dagger \phi_2 \right)
+ \mathrm{H.c.} \right]
\label{seminal}
\ee
of ref.~\cite{geometricT}.
In this case
\be
D = \left( \begin{array}{ccc}
2 & -1 & -1 \\ -1 & 2 & -1 \\ -1 & -1 & 2
\end{array} \right)
\ee
and
\be
\bar D = \left( \begin{array}{cc}
2 & -1 \\ -1 & 2
\end{array} \right).
\ee
Therefore,
according to eq.~(\ref{jbhur}),
a particular spontaneously-$CP$-breaking solution with geometrical phases
is $\theta_2 = \theta_3 = \psi_0$ while $\theta_1 = 0$.
The value of $\psi_0$ is given by eq.~(\ref{psisol}):
$\psi_0 = \pm 2 \pi / 3$ for the two $CP$-conjugate solutions.

Notice that in this case the vector $\eta$ of eq.~(\ref{very-particular})
is \emph{not} a solution to eqs.~(\ref{eqn-theta-i}),
since
\be
\frac{2 \pi}{3}\, \bar D \left( \begin{array}{c} 1 \\ 2 \end{array} \right) =
\left( \begin{array}{c} 0 \\ 2 \pi \end{array} \right)
\ee
does not coincide with $\pm \left( 2 \pi / 3 \right) \left( 1,\, 1 \right)^T$
as one would wish.

The general solution involves the solution to eq.~(\ref{homogeneous}),
which is identical with the solutions for the rephasing symmetry group.
Since the potential~(\ref{seminal}) is of the form $V_{213}$ with $N=3$,
the symmetry group is $\Z_{\left| s_4 \right|} = \Z_3$ with
$\phi_2 \to \omega \phi_2, \ \phi_3 \to \omega^{-1} \phi_3$.
Therefore,
the geometrically-$CP$-breaking solutions are
\begin{subequations}
\label{sol3}
\bea
\left( \theta_1,\ \theta_2,\ \theta_3 \right) &=&
\left( 0,\ \frac{2 \pi}{3},\ \frac{2 \pi}{3} \right),
\label{jviwq} \\
\left( \theta_1,\ \theta_2,\ \theta_3 \right) &=&
\left( 0,\ \frac{4 \pi}{3},\ 0 \right),
\\
\left( \theta_1,\ \theta_2,\ \theta_3 \right) &=&
\left( 0,\ 0,\ \frac{4 \pi}{3} \right),
\eea
\end{subequations}
and their $CP$ conjugates.
The solution~(\ref{jviwq}) is equivalent,
through a rephasing of the doublets,
to $\left( \theta_1,\ \theta_2,\ \theta_3 \right)
= \left( - 2 \pi / 3,\ 0,\ 0 \right)
= \left( 4 \pi / 3,\ 0,\ 0 \right)$,
which renders evident the symmetry among the three solutions.
To the solutions~(\ref{sol3}) must of course be added their $CP$ conjugates.

\subsection{Second example: $V_{213}$ for $N=5$}
\label{sec:pgfity}

In this case the matrix $D$ and its Smith normal form are
\be
\left(\begin{array}{ccccc}
2 & -1 & -1 & 0 & 0 \\ 
0 & 2 & -1 & -1 & 0\\
0 & 0 & 2 & -1 & -1 \\ 
-1 & 0 & 0 & 2 & -1 \\
-1 & -1 & 0  & 0 & 2  
\end{array}\right)
\quad \mbox{and} \quad
\left(\begin{array}{ccccc}
1 & 0 & 0 & 0 & 0 \\ 
0 & 1 & 0 & 0 & 0\\
0 & 0 & 1 & 0 & 0 \\ 
0 & 0 & 0 & 11 & 0 \\
0 & 0 & 0  & 0 & 0  
\end{array} \right),
\label{aaav213}
\ee
respectively.
The symmetry group is therefore $\Z_{11} = \Z_{\left| s_6 \right|}$ which,
according to section~\ref{subsection-certain-potentials}, 
is generated by phase rotations through
\be
\sigma_5 = \frac{2 \pi}{11} \left(
0,\, 1,\, 10,\, 3,\, 6 \right).
\ee
To keep the notation short,
we write here and below not the phase transformation themselves,
but only their phase shifts;
then,
instead of taking powers of the transformations,
we must write multiples of the phase rotations.
The value of $\psi_0$ is $\psi_0 =  \pi \pm \pi/5 = \pm 4 \pi / 5$.
Since
\be
\bar D = \left(\begin{array}{cccc}
2 & -1 & -1 & 0\\
0 & 2 & -1 & -1 \\ 
0 & 0 & 2 & -1 \\
-1 & 0  & 0 & 2  
\end{array}\right),
\ee
the particular values of the $\theta_k$ of the form (\ref{very-particular}),
\be
\eta_5 = \frac{2 \pi}{5} \left( 0, 1, 2, 3, 4 \right)
\ee
indeed provide a particular solution to eq.~(\ref{eqn-theta-i}),
since one obtains $\psi_2 = \psi_3 = \psi_4 = \psi_5 = 4 \pi / 5$.
Therefore,
the generic calculable phases of the VEVs are in this case
\be
\theta^{(q)} = \eta_5 + q \sigma_5
= \frac{2 \pi}{55} \left[ \left( 0,\ 11,\ 22,\ 33,\ 44 \right)
+  q \left( 0,\, 5,\, 50,\, 15,\, 30 \right) \right],
\label{theta5}
\ee
for $q = 0, 1, \ldots, 10$.
Including their $CP$ conjugates,
there are 22 distinct degenerate global minima with calculable phases,
related among themselves by rephasing symmetry transformations.

Let us stress that any cyclic permutation of the phases in eq.~(\ref{theta5})
also leads to viable minima.
These are {\em not}\/ additional minima,
since they are already included in eq.~(\ref{theta5}).
Indeed,
we remind the reader that all the phases are defined
up to overall phase rotations---this is the gauge symmetry $U(1)$
discussed at the end of section~\ref{subsection-smith}.
We always set $\theta_1 =0$
in order to pick a representative point
of each family of relative phase configurations.
Thus,
a cyclic permutation of $\eta_5$ is equivalent to $\eta_5$.
Moreover,
a cyclic permutation of $\sigma_5$
may be expressed as a multiple of $\sigma_5$;
for instance,
if $b$ is the generator of the cyclic permutation group $\Z_5$
defined by the transformation~(\ref{cyclic}),
then
\be
b^{-1}\sigma_5 b =
\frac{2 \pi}{11} \left(6,\, 0,\, 1,\, 10,\, 3\right)
\equiv
\frac{2 \pi}{11} \left(0,\, 5,\, 6,\, 4,\, 8\right)
= 5 \sigma_5.
\label{closure}
\ee
Therefore,
a cyclic permutation of any $\theta^{(q)}$ in eq.~(\ref{theta5})
just yields $\theta^{(5q)}$.

Let us also mention that,
because of eq.~(\ref{closure}),
the symmetry group of the potential
is the semi-direct product $\Z_{11}\rtimes \Z_5$,
plus the gauge-$U(1)$ symmetry group of overall phase rotations.

\subsection{Other potentials for $N=5$}

In section~\ref{section-stable} we have found
four distinctly looking potentials,
all of them protected by stable rephasing symmetries
of the $\Z_{11}$ type.
They were generically written $V_{1+k,\, 1,\, 1+2k}$ for $k=1,2,3,4$.
In subsection~\ref{sec:pgfity} we have analyzed the case $k=1$,
\textit{i.e.}\ the potential $V_{213}$.
Geometrical $CP$ violation for the remaining three potentials
can be easily written by using the same parameter $k$.
Namely,
for $N=5$,
the potential $V_{1+k,\, 1,\, 1+2k}$ has a particular solution $p \eta_5$,
where $p=1$ for $k=1$ and $k=4$ and $p=2$ for $k=2$ and $k=3$.
In each case, adding to this particular solution
an arbitrary transformation from the corresponding rephasing symmetry group
generated by (\ref{sym5Vk}) 
yields the calculable phases of all possible global minima.
We thus conclude that {\em any} potential protected with a stable symmetry
for $N=5$ can exhibit geometrical $CP$ breaking,
with the VEVs acquiring calculable phases.

Noticing that $2\eta_5$ is essentially equivalent to $\eta_5$
up to a permutation of the doublets,
we conclude that in all cases
{\em the VEVs exhibiting geometrical $CP$ violation for $N=5$
represent the vertices of the regular pentagon
on the complex plane.}\footnote{The relevance of the regular pentagon
to geometrical $CP$ violation with $N=5$
was also noted in ref.~\cite{geometricCP}.}

\subsection{Third example: $V_{213}$ for $N=7$}

The symmetry group is $\Z_{43}$ with generator
\be
\sigma_7 = \frac{2 \pi}{43} \left( 0,\, 1,\, 42,\, 3,\, 38,\, 11,\, 22 \right). 
\ee
It can be verified again that 
\be
\eta_7 = \frac{2 \pi}{7} \left( 0,\, 1,\, 2,\, 3,\, 4,\, 5,\, 6 \right)
\label{eta7}
\ee
is a particular solution,
so that a generic global minimum has calculable phases
\be 
\theta^{(q)} = \eta_7 + q \sigma_7
\quad \mbox{for} \ q = 0, 1, \ldots, 42.
\label{theta7}
\ee
For the other potentials protected by the stable symmetry,
listed in~(\ref{stable7}) as $V_{1+k,\, 1,\, 1+2k}$,
we have the same situation as in the case $N=5$,
namely,
in each case a particular solution is $p \eta_7$
with $p=1$ for $k=1,6$,
$p=3$ for $k=2,5$,
and $p=2$ for $k=3,4$. 
However,
all these seemingly disparate cases are in fact,
up to a permutation of the doublets,
of the same form as $\eta_7$.

\subsection{Fourth example: $V_{214}$ for $N=7$}

We now have
\be
\bar D = \left( \begin{array}{cccccc}
2 & -1 & 0 & -1 & 0 & 0 \\
0 & 2 & -1 & 0 & -1 & 0 \\
0 & 0 & 2 & -1 & 0 & -1 \\
0 & 0 & 0 & 2 & -1 & 0 \\
-1 & 0 & 0 & 0 & 2 & -1 \\
0 & -1 & 0 & 0 & 0 & 2
\end{array} \right).
\ee
Therefore,
$\bar D \left( 1,\, 2,\, 3,\, 4,\, 5,\, 6\right)^T
= \left( -4,\, -4,\, -4,\, 3,\, 3,\, 10 \right)^T
= \left( 3,\, 3,\, 3,\, 3,\, 3,\, 3 \right)^T$ modulo 7.
Thus,
$\eta_7$ is once again a particular solution.
The general solution is
\be
\left( \theta_1,\, \theta_2,\, \theta_3,\, \theta_4,\, \theta_5,\,
\theta_6,\, \theta_7 \right)
= \eta_7 + \frac{2 \pi q}{29} \left( 0,\, 1,\, 8,\, 28,\, 23,\, 17,\, 4 \right)
\ee
according to eq.~(\ref{rho}).
Once again,
there are further cases $V_{1+k,\, 1,\, 1+3k}$ for $k = 2, \ldots, 6$,
but it turns out that they are all identical to the case $V_{214}$
up to permutations of the doublets.
Therefore,
{\em all the cases with VEVs exhibiting geometrical $CP$ breaking for $N=7$
represent the vertices of the regular heptagon on the complex plane}.

\section{Beyond stable symmetries for even $N$}\label{section-beyond}

In section~\ref{subsection-calculating-phases} we have shown that
in the case of an even number of doublets
there is no room for geometrical $CP$ violation
when the potential is protected by a stable symmetry.
One may ask whether this conclusion changes
when the requirement of stability is removed.
In the following we present an argument against this possibility.

Let us firstly consider the simple case $N=4$. 
As we have proved in section~\ref{section-stable},
the potentials $V_{213}$ and $V_{234}$ are then invariant
under the same rephasing group $\Z_5$.
Suppose then that both terms are present in the potential:
\bea
V_{\rm phase} &=& \lambda \left[
\left( \fdf{1}{2} \right) \left( \fdf{1}{3} \right) + {\rm cyclic} +
{\rm H.c.}\right]\nonumber\\
& & + \lambda'
\left[
\left( \fdf{1}{2} \right) \left( \fdf{3}{4} \right) + 
\left( \fdf{2}{3} \right) \left( \fdf{4}{1} \right) + 
{\rm H.c.}\right].
\eea
Using the \textit{Ansatz}\/
$\left\langle \phi_k^0 \right\rangle_0 = u e^{i\theta_k}$,
one obtains
\be
\left\langle V_{\rm phase} \right\rangle_0 = 2 u^4
\left( \lambda J + \lambda' J' \right),
\ee
where
\begin{subequations}
\bea
J &=& \cos{\psi_1} + \cos{\psi_2} + \cos{\psi_3} + \cos{\psi_4},
\\
J' &=& \cos{\left( \psi_1 + \psi_3 \right)}
+ \cos{\left( \psi_2 + \psi_4 \right)}.
\eea
\end{subequations}
If only the first term with positive $\lambda$ was present,
there would be no spontaneous $CP$ violation,
because the minimum would be given,
up to a rephasing symmetry,
by phases $\left( \theta_1,\, \theta_2,\, \theta_3,\, \theta_4 \right)
= \left( 0,\, \pi,\, 0,\, \pi \right)$.
The question is whether the presence of the second term may lead to non-trivial,
calculable phases.

To answer this question we have used
the geometrical minimization methods
described in ref.~\cite{minimization2012}.
Namely,
we have considered the $\left(J,\, J' \right)$ plane,
drawn the region covered by all possible phase configurations,
and inspected it for the presence of sharp corners.
One may find the shape of the allowed region
analytically:\footnote{Equation~(\ref{uvity}) may be derived
in the following way.
Consider four planar vectors $\vec n_1$,
$\vec n_2$,
$\vec n_3,$
and $\vec n_4$,
all of them with unit length and with polar angles $\psi_1$,
$\psi_2$,
$-\psi_3$,
and $-\psi_4$,
respectively.
We define two vectors $\vec p = \vec n_1 + \vec n_3$
and $\vec q = \vec n_2 + \vec n_4$.
Then,
\[
4+2J' =
2 + 2 \cos{\left( \psi_1 + \psi_3 \right)}
+ 2 + 2 \cos{\left( \psi_2 + \psi_4 \right)}
= \left| \vec n_1 + \vec n_3 \right|^2 + \left| \vec n_2 + \vec n_4 \right|^2
= \left| \vec p \right|^2 + \left| \vec q \right|^2.
\]
On the other hand,
$J$ is the projection of $\vec p + \vec q$ on the horizontal axis.
Then,
\[
J^2 \le \left| \vec p + \vec q \right|^2
= 2 \left( \left| \vec p \right|^2 + \left| \vec q \right|^2 \right)
- \left| \vec p - \vec q \right|^2
\le 2 \left( \left| \vec p \right|^2 + \left| \vec q \right|^2 \right)
= 8 + 4J'.
\]}
\begin{subequations}
\bea
|J'| &\le& 2,
\\
J^2 &\le& 4 J' + 8.
\label{uvity}
\eea
\end{subequations}
This region has only two sharp vertices,
$\left( J,\, J' \right) = \left( \pm 4,\, 2 \right)$.
Both of them correspond to $CP$-conserving alignments:
$\left( \theta_1,\, \theta_2,\, \theta_3,\, \theta_4 \right)
= \left( 0,\, 0,\, 0,\, 0 \right)$
for $\left( J,\, J' \right) = \left( 4,\, 2 \right)$
and $\left( \theta_1,\, \theta_2,\, \theta_3,\, \theta_4 \right)
= \left( 0,\, \pi,\, 0,\, \pi \right)$
for $\left( J,\, J' \right) = \left( -4,\, 2 \right)$.
All other alignments are not calculable.
This proves the absence the geometrical $CP$ violation in this model.

The above arguments may be extended to a generic situation with even $N$.
We start with some set of terms $V_{klm}$
and define the corresponding $J = \sum_{j=1}^N \cos{\psi_j}$. 
The fact that $V_{klm}$ is not protected by a stable symmetry means that 
it is possible to write a second set of terms,
which leads to a $J'$ containing---at least for small $N$---the cosines
of pairs of $\psi_j$'s.
One then derives the allowed region in the $\left( J,\, J' \right)$ plane:
\begin{subequations}
\bea
|J'| &\le& \frac{N}{2},
\\
J^2 &\le& N J' + \frac{N^2}{2},
\eea
\end{subequations}
which has the same geometry as before
and has only two $CP$-conserving corners.

This discussion suggests the conclusion that for an even number of doublets
it is much more difficult---maybe impossible---to achieve
geometrical $CP$ violation with renormalizable potentials.
Then the only way to do it would be to include higher order
(non-renormalizable)
terms,
as discussed in ref.~\cite{geometricCP}.

\section{Conclusions}

Since the fundamental origin of $CP$ violation is still unknown,
any theoretical framework in which $CP$ violation arises naturally
is worth a dedicated study.
Geometrical $CP$ violation,
as suggested in ref.~\cite{geometricT},
is one such framework.
The idea of calculable $CP$-violating phases
which spontaneously arise in the vacuum of an extended Higgs sector
is rather general and goes beyond the particular 3HDM example
presented in ref.~\cite{geometricT}.
In this paper we have investigated this phenomenon
in the context of models with more than three Higgs doublets.
We have shown how one may find
the rephasing symmetry group of any cyclic $N$-Higgs-doublet potential
and have given several examples of geometrical $CP$ violation
for $N=5$ and $N=7$. 
In all those cases we have found a VEV alignment in the form
of a regular polygon in the complex plane,
distinct from what happens in the $N=3$ model of ref.~\cite{geometricT}.
In contrast,
for an even number of doublets we have proved the impossibility
of geometrical $CP$ violation for phase-sensitive potentials
which are protected by a `stable' symmetry
and have given an argument that suggests that
it might be impossible to achieve geometrical $CP$ violation
even for more general renormalizable potentials.

\vspace*{5mm}

\section*{Acknowledgements} 
The work of I.P.I.\ was supported in part by the grant RFBR 11-02-00242-a,
the RF President Grant for scientific schools NSc-3802.2012.2,
and the Program of the Department of Physics
of the Scientific Council of the Russian Academy of Sciences
``Studies of Higgs boson and exotic particles at LHC.''
The work of L.L.\ was supported by Portuguese national funds
through the project PEst-OE/FIS/UI0777/2011
of the \textit{Funda\c c\~ao para a Ci\^encia e a Tecnologia}.

\end{document}